\documentclass{wscpaperproc}
\usepackage{latexsym}
\usepackage{graphicx}
\usepackage{subfigure}
\usepackage{mathptmx}
\usepackage[T1]{fontenc}

\usepackage{amsmath}
\usepackage{amsfonts}
\usepackage{amssymb}
\usepackage{amsbsy}
\usepackage{amsthm}

\usepackage[pdftex,colorlinks=true,urlcolor=blue,citecolor=black,anchorcolor=black,linkcolor=black]{hyperref}

\newtheoremstyle{wsc}
{3pt}
{3pt}
{}
{}
{\bf}
{}
{.5em}
{}

\theoremstyle{wsc}

\begin{document}

%
%

\pagestyle{fancyplain}

\thispagestyle{plain}
\firstPageHead{}

\chead{\fancyplain{}{\itshape Salazar-Serna, Cadavid, and Franco}}

\rhead{}
\cfoot{}
\renewcommand{\headrulewidth}{0pt} 

\makeatletter
\let\@internalcite\cite
\def\cite{\def\@citeseppen{-1000}%
    \def\@cite##1##2{(##1\if@tempswa , ##2\fi)}%
    \def\citeauthoryear##1##2##3{##1 ##3}\@internalcite}
\def\citeNP{\def\@citeseppen{-1000}%
    \def\@cite##1##2{##1\if@tempswa , ##2\fi}%
    \def\citeauthoryear##1##2##3{##1 ##3}\@internalcite}
\def\citeN{\def\@citeseppen{-1000}%
    \def\@cite##1##2{##1\if@tempswa, ##2)\else{}\fi}%
    \def\citeauthoryear##1##2##3{##1 (##3)}\@citedata}
\def\citeA{\def\@citeseppen{-1000}%
    \def\@cite##1##2{(##1\if@tempswa , ##2\fi)}%
    \def\citeauthoryear##1##2##3{##1}\@internalcite}
\def\citeANP{\def\@citeseppen{-1000}%
    \def\@cite##1##2{##1\if@tempswa , ##2\fi}%
    \def\citeauthoryear##1##2##3{##1}\@internalcite}
\def\shortcite{\def\@citeseppen{-1000}%
    \def\@cite##1##2{(##1\if@tempswa , ##2\fi)}%
    \def\citeauthoryear##1##2##3{##2 ##3}\@internalcite}
\def\shortciteNP{\def\@citeseppen{-1000}%
    \def\@cite##1##2{##1\if@tempswa , ##2\fi}%
    \def\citeauthoryear##1##2##3{##2 ##3}\@internalcite}
\def\shortciteN{\def\@citeseppen{-1000}%
    \def\@cite##1##2{##1\if@tempswa, ##2\else{}\fi}%
    \def\citeauthoryear##1##2##3{##2 (##3)}\@citedata}
\def\shortciteA{\def\@citeseppen{-1000}%
    \def\@cite##1##2{(##1\if@tempswa , ##2\fi)}%
    \def\citeauthoryear##1##2##3{##2}\@internalcite}
\def\shortciteANP{\def\@citeseppen{-1000}%
    \def\@cite##1##2{##1\if@tempswa , ##2\fi}%
    \def\citeauthoryear##1##2##3{##2}\@internalcite}
\def\citeyear{\def\@citeseppen{-1000}%
    \def\@cite##1##2{(##1\if@tempswa , ##2\fi)}%
    \def\citeauthoryear##1##2##3{##3}\@citedata}
\def\citeyearNP{\def\@citeseppen{-1000}%
    \def\@cite##1##2{##1\if@tempswa , ##2\fi}%
    \def\citeauthoryear##1##2##3{##3}\@citedata}
%
%
%
\def\@citedata{%
    \@ifnextchar [{\@tempswatrue\@citedatax}%
                  {\@tempswafalse\@citedatax[]}%
}

\def\@citedatax[#1]#2{%
\if@filesw\immediate\write\@auxout{\string\citation{#2}}\fi%
  \def\@citea{}\@cite{\@for\@citeb:=#2\do%
    {\@citea\def\@citea{, }\@ifundefined
       {b@\@citeb}{{\bf ?}%
       \@warning{Citation `\@citeb' on page \thepage \space undefined}}%
{\csname b@\@citeb\endcsname}}}{#1}}%

%
\def\@citex[#1]#2{%
\if@filesw\immediate\write\@auxout{\string\citation{#2}}\fi%
  \def\@citea{}\@cite{\@for\@citeb:=#2\do%
    {\@citea\def\@citea{; }\@ifundefined
       {b@\@citeb}{{\bf ?}%
       \@warning{Citation `\@citeb' on page \thepage \space undefined}}%
{\csname b@\@citeb\endcsname}}}{#1}}%

%
\def\@biblabel#1{}
\makeatother



\newdimen\bibindent
\bibindent=0.0em
\def\thebibliography#1{\section*{\refname}\list
   {}{\settowidth\labelwidth{[#1]}
   \leftmargin\parindent
   \itemindent -\parindent
   \listparindent \itemindent
   \itemsep 0pt
   \parsep 0pt}
   \def\newblock{}
   \sloppy
   \sfcode`\.=1000\relax}


\setlength{\baselineskip}{12.7pt}

\title{MODELING URBAN TRANSPORT CHOICES: INCORPORATING SOCIOCULTURAL ASPECTS}

\author{\begin{center}Kathleen Salazar-Serna\textsuperscript{1,2}, Lorena Cadavid\textsuperscript{2}, and Carlos J. Franco\textsuperscript{2}\\
[11pt]
\textsuperscript{1}Dept.~of Civil and Industrial Eng., Pontificia Universidad Javeriana, Cali, VA, COLOMBIA\\
\textsuperscript{2}Dept.~of Computing and Decision Sciences, Universidad Nacional de Colombia, Medellín, ANT, COLOMBIA\\
\end{center}
}

\maketitle

\vspace{-12pt}

\section*{ABSTRACT}
This paper introduces an agent-based simulation model aimed at understanding urban commuters' mode choices and evaluating the impacts of transport policies to promote sustainable mobility. Crafted for developing countries, where utilitarian travel heavily relies on motorcycles, the model integrates sociocultural factors that influence transport behavior. Multinomial models and inferential statistics applied to survey data from Cali, Colombia, inform the model, revealing significant influences of sociodemographic factors and travel attributes on mode choice. Findings highlight the importance of cost, time, safety, comfort, and personal security, with disparities across socioeconomic groups. Policy simulations demonstrate positive responses to interventions like free public transportation, increased bus frequency, and enhanced security, yet with modest shifts in mode choice. Multifaceted policy approaches are deemed more effective, addressing diverse user preferences. Outputs can be extended to cities with similar sociocultural characteristics and transport dynamics. The methodology applied in this work can be replicated for other territories.

\section{INTRODUCTION}
\label{sec:intro}

The analysis of transport mode choices is crucial for urban planners and policy makers. By understanding how users decide on their commuting modes, it is possible to identify factors that can be influenced to change travel behavior and promote the adoption of more sustainable transportation modes. Agent-based modeling (ABM) is particularly valuable for this purpose, as it can represent complex systems like transportation and identify emerging collective behaviors resulting from the autonomous decisions of transport users interacting among them and with the environment \cite{kagho2020agent}. These capabilities make ABM suitable for analyzing the impacts of transport policies \cite{wise2017transportation}. However, the application of ABM in analyzing transport mode choices has been limited and studies have been conducted predominantly in developed countries \cite{cadavid2021mapping,salazar2023simulating}. The effectiveness of these findings may not extend seamlessly to developing regions due to different contextual policy needs and the distinct ways socioeconomic and cultural factors influence human behavior \shortcite{carley1991theory,salazar2023social}. Therefore, policies that have been successful in one setting might not achieve similar outcomes in another. 
\\

Previous studies in transportation have identified various determinants affecting mode choice. These factors can be grouped into several categories: sociodemographic characteristics such as age, sex, occupation, and income level \shortcite{ashalatha2013mode}; travel habits including distance traveled, travel time, origin-destination pairs, and trip purpose \shortcite{madhuwanthi2016factors}; and attributes of the built environment like design, density, and capacity \cite{Ewing2010}. Additionally, attitudes and perceptions regarding transport modes, which cover aspects such as comfort, cost, security, safety, quality, and reliability, play a crucial role \cite{FuXumei}. Culture, as Olson suggests, profoundly shapes these attitudes by influencing beliefs and expectations about transport modes \cite{olson2003culture}.       
\\

Models are designed to simulate the behavior of real systems. When representing human behavior, it is crucial to include the sociocultural factors that influence such behaviors. Although prior research offers insights into these factors, accurately defining them based on the models developed for a different context can be challenging.  In the global South, the dynamics of utilitarian transportation heavily rely on two- and three-wheeler vehicles with engine capacities below 200 cc, predominantly owned by low- and middle-income families \cite{PAHO}. These vehicles are favored for their flexibility, low cost, and the ability to provide quick commutes—an essential alternative to often inadequate public transportation systems \cite{chiu2023,tanabe2018}. In contrast, in developed countries, motorcycles, particularly those designed for adventure and touring, are primarily used for leisure \cite{broughton}. The substantial and increasing prevalence of motorcycles in developing countries presents additional challenges for policymakers in terms of emission control, road fatalities, and managing extended travel times caused by congestion \cite{bakker2018,suatmadi2019}. Given these complex and particular conditions, there is a pressing need for models that are specifically tailored to the sociocultural contexts of these regions.          
\\

This paper presents the results from an agent-based simulation model designed to represent utilitarian travelers in developing countries. These travelers interact within the model to make decisions about their primary means of transportation. In an effort to provide a tool to analyze the impacts of transport policy, we developed a custom model that not only incorporates motorcycles as the primary mode, but also integrates key sociocultural factors influencing travel behavior. We gathered these sociocultural insights through focus groups and a designed survey conducted in Cali, Colombia, a city representative of the typical features and dynamics found in major cities across developing nations. To analyze the survey data, multinomial models were implemented and statistical tests were used to assess differences in perceptions across socioeconomic groups. These insights were used to inform the model by defining variables and parameters and setting them up for the case study city.  
\\

In the following sections, we describe the methodology used to identify the influential factors of mode choice represented in our model and detail the process for calculating key parameters that represent sociocultural factors. Finally, we present the outcomes of a base case scenario compared with three proposed policies aimed at encouraging the use of public transit and analyze their impacts, considering sociocultural variances.          
\\


\section{METHODS}

\subsection{Agent-based simulation model}

Initially, a high-level conceptual model was established, grounded in previous research on mode choices using ABM and integrating consumer behavior principles that drive individuals' decision-making \shortcite{Faboya2020,kangur2017agent}. Inspired in the CONSUMAT approach \cite{jager2012updated}, this model accounts for various factors to apply the cognitive process, including personal trip satisfaction and uncertainty about trying new transportation modes, and emphasizes the role of social interactions; agents gather information and make comparisons within their social network to deal with the uncertainty and make decisions. 
\\

\begin{figure}[htb]
{
\centering
\includegraphics[width=0.4\textwidth]{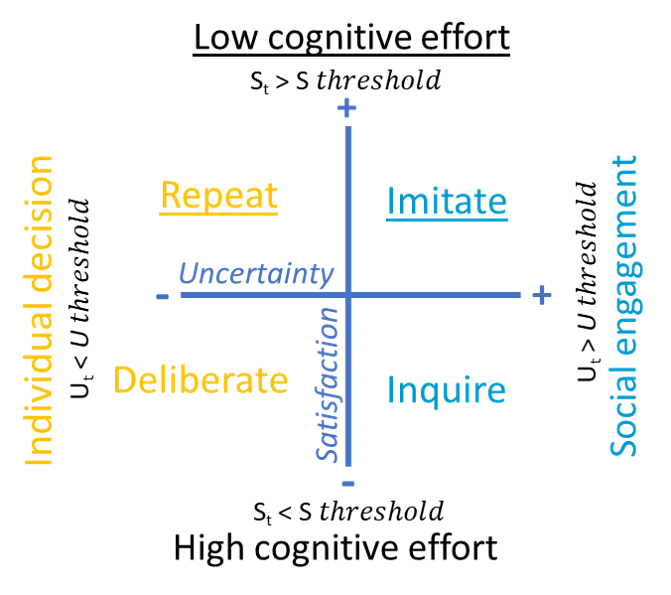}
\caption{Strategies followed by agents to make decisions.  \mbox{$S_t =$} Satisfaction at time step t, \mbox{$U_t =$} Uncertainty at time step t}  \label{fig:consum}
}
\end{figure}

Agents begin the simulation using a mode of transportation among private car, motorcycle, or public transit. The initial distribution of transport users depends on their sociodemographic characteristics. Throughout the simulation, modal shifts take place during decision periods based on the strategies agents adopt after assessing their mental models. Fig. \ref{fig:consum} illustrates the cognitive process that agents follow to select one of the decision-making strategies.        

Trip satisfaction is calculated as a weighted sum of satisfaction levels for each factor affecting mode choice. The subsequent subsections describe the factors considered in the satisfaction function. The uncertainty arises from insufficient knowledge about transport modes, which can be mitigated by combining personal experience using them with peer experiences from one's social network; two parameters representing uncertainty avoidance and collective belongingness weight the uncertainty level. The resulting satisfaction and uncertainty values are then compared to individual thresholds for desired satisfaction and tolerable uncertainty; based on this comparison, agents repeat, imitate, inquire, or deliberate to select a commuting mode for the subsequent period. Those who imitate the most used mode or inquire (compare the possible satisfaction with others' modes) interact with their social network to make decisions. A preferential attachment topology connects agents in the model, with a higher likelihood of forming ties with similar individuals within the same socioeconomic group. See \cite{salazar2023social} for more information of the social network impact on the mode choice. 
\\

\begin{figure}[htb]
{
\centering
\includegraphics[width=0.95\textwidth]{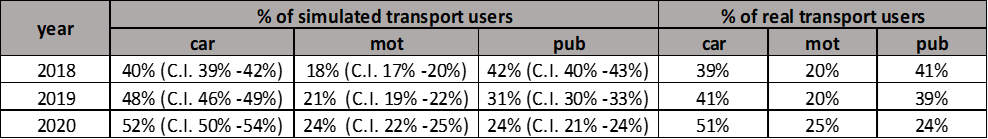}
\caption{Comparison of transport user distribution: ABM simulation vs. real data}  \label{fig:valid}
}
\end{figure}

\begin{figure}[htb]
{
\centering
\includegraphics[width=0.8\textwidth]{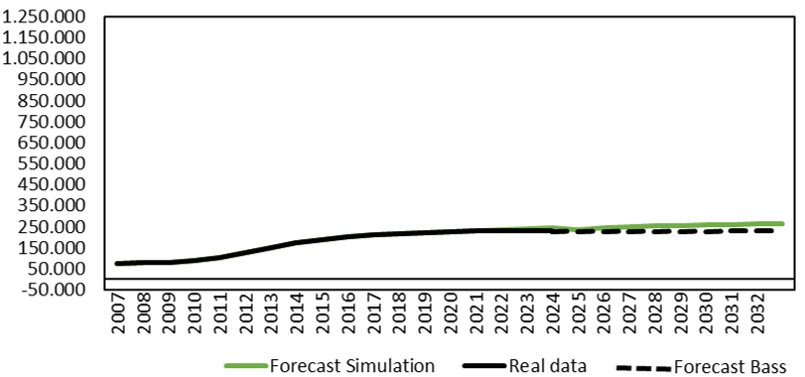}
\caption{Comparison of transport user distribution: Bass model forecast vs. ABM simulation}  \label{fig:bass}
}
\end{figure}

The model was validated using the validation in parts technique \cite{carley1996validating}, where inputs, processes, and outputs are incrementally validated. Consistency was evaluated using extreme values, and outputs were assessed using data from Cali city. To evaluate the outputs, the model was initialized and run 100 times for three periods with distributions corresponding to 2018. The proportions of transport users were then compared with the real data in 2020 \cite{cali}. As shown in Figure \ref{fig:valid}, the simulation results replicate the emerging behavior where private alternatives increase the proportion of users, and the public service decreases. A second analysis was carried out comparing the 10-year forecast of the simulation against the S-curve calculated with the Bass model of diffusion for motorcycles \cite{coronado2018application}. The nonlinear Bass model was implemented in R software using motorcycle registries in Cali between 2007 and 2023. Results obtained with the agent-based model follow a similar trajectory for the number of new motorcycles expected to be in circulation in the next decade, according to the Bass forecast (see Figure \ref{fig:bass}).

\subsection{Adapting the simulation model to the context and finding information to parameterize it}

After developing the decision rules and submodels for the simulation, we defined the necessary information to initialize the model with Cali as the reference city, capturing the main characteristics of transport dynamics typical in developing countries. Parameters such as road accident rates, the initial distribution of users by transport mode, and the wealth, gender, and age distribution of the synthetic population were determined using historical records from Cali \cite{cali}. Parameters associated with travel behavior are generally available in origin-destination surveys; however, the most recent update for this survey in Cali was in 2015 \cite{encuesta}, and since then, urban development has significantly altered the city's mobility dynamics. Moreover, there is a group of parameters related to the sociocultural characteristics of agents that influence decision-making, such as perceptions of transport attributes (travel cost, road safety, personal security, comfort, air pollution), uncertainty avoidance, and collectivism levels, are often unmeasured or omitted in household or origin-destination surveys. Given this gap in the available information and the need to accurately represent the current context of the city under study, it was decided to enhance our research by designing and implementing a new survey. 
\\

The survey was meticulously crafted, drawing from both established determinants of travel mode choices identified in existing literature \shortcite{agarwal2020calibration,ahmed2021understanding,fan2020exploring,gadepalli2020role} and insights garnered from two focus groups involving transport users from Cali, Colombia. Structured to gather comprehensive data, the questionnaire delves into respondents' sociodemographic profiles, travel habits, and perceptions regarding public transit, private cars, and motorcycles. For further details, the supplementary material provides access to the survey dictionary. A dashboard that offers an interactive summary of the survey results can be found in \citeN{survey}.
\\

\subsection{Identifying the factors influencing the mode choice}
The factors affecting transport mode choices, as recognized in previous studies, were compared with insights from the focus groups and the survey carried out in Cali to establish their incorporation into the simulation model. For example, personal security, defined as the likelihood of being robbed, assaulted, or harassed, surfaced as a crucial aspect in trip satisfaction that has often been overlooked in research conducted in developed nations but holds great significance in developing regions. The statistical significance of personal security and other factors considered for the mode satisfaction was confirmed using the data collected with the survey.
\\

To conduct the survey, a sample of 385 individuals was determined using simple random sampling without replacement in a population of 2.2 million. This sample size was calculated by applying Cochran's formula with a 95\% confidence level and a margin of error of ±5\%, assuming maximum estimated variance. Data were collected from 970 voluntary respondents who were urban commuters in Cali city. Categorical variables are summarized by counts and proportions, while continuous variables are reported with the mean or median depending on whether they are normally distributed or not. Discrete choice models, founded on random utility theory, have been a widely accepted approach for analyzing determinants in travel mode choice for decades, with the multinomial logit model being predominant due to its simplicity and explicability \cite{Domencic}. Using multinomial logit (MNL) and probit (MNP) models, the determinants of mode choice were empirically identified using the Wald test with the survey data. These results were considered in the development and parameterization stages of the simulation model; the factors identified as influential in the decision making were included in the model as agents' attributes or global variables.            
\\

To accurately represent the agents' attributes in the simulation model, capturing the sociocultural aspects influencing travel behavior is essential. One of the purposes of the survey data analysis is to establish whether there are differences among the various commuter groups regarding their perceptions and the importance they assign to factors affecting mode choice. Comparisons of groups by socioeconomic status and transport mode were performed using a Kruskal-Wallis test followed by a Mann-Whitney U test for pairwise comparisons with continuous variables. Categorical variables were compared using the Chi-square ($\chi^2$) test of independence or Fisher's exact test.

\subsection{Simulation experiments}
The model implemented in NetLogo 6.4.0 was initialized with the values for Cali city corresponding to 2022. A base case scenario was run 80 times after checking the number of repetitions at which the coefficient of variation in results stabilizes. Time steps are counted as ticks that are equivalent to 2 minutes of a peak hour on a weekday. During 30 ticks, agents accumulate data from the system and their own commute to evaluate their mental model and implement a strategy to make a decision about their transport mode. This time period represents a typical commute year that constitutes the decision period for the agents. We simulated 10 years to analyze how the distribution of transport users changes over time. In addition, indicators such as average travel time, speed, CO2 emissions, and accident rates are monitored in the system.      
\\

The base case scenario exhibits the general behavior of the actual system showing a decreasing participation of public users who are migrating to private options (cars and motorcycles); with no interventions in the system, this trend is expected to continue for the coming years. As the ultimate purpose of the model is to serve as a test bed to evaluate the effectiveness of public transport policies for more sustainable mobility, a second experiment was run to analyze the impacts of three policies oriented to increase the public transit usage. According to the survey, one of the most valuable factors for users is travel time, and the public service has both long travel times and long waiting times; It is expected that reducing the time to take the bus could have a positive impact on public transit users, influencing their behavior to continue using this service. Cost of travel and personal security are also top concerns for users, (section \ref{sec:3.2} presents ratings users give to the transport characteristics). Therefore, increased service frequency, fare-free and personal security as public transport policies were individually evaluated. A third group of experiments was analyzed after introducing simultaneous interventions with combinations of the previous policies. 
\\


\section{RESULTS AND DISCUSSION}

\subsection{Influencing factors identified with the discrete choice models}

After running the MNL and MNP models, identical qualitative results were obtained. The supplementary material contains a link to a folder with the Stata output of the models and a table summarizing the individual significance of variables. The Wald test shows that mode choice is significantly influenced by: (1) sociodemographic and economic conditions, including gender, age, and socioeconomic status; (2) travel behavior aspects such as travel time; and (3) transport attributes evaluated by travelers, including cost, time, comfort, road safety, personal security, and pollution. These findings align with previous research on commute mode choices. Studies have identified a gender gap in vehicle ownership \cite{Havet2021}. The results from the multinomial models suggest women are more likely to use public transportation than either a car or motorcycle, a trend that may be explained by economic differences and distinct travel habits compared to men \cite{Rosenbloom2006}.
\\

In our model, the probabilities of starting the simulation with a car or motorcycle are higher for men, while the assignment rules give a lower probability of having private transportation to younger people, particularly those in middle and low income brackets. Previous research has shown that as people age, their use of public transportation decreases \shortcite{Basaric2016}, and as economic power increases, so does the desire for the status associated with owning a car or motorcycle \cite{CAF2015}.
\\

Multinomial models also reveal that comfort and personal security are significant factors in choosing a car, whereas cost concerns tend to sway users towards motorcycles. Specifically, in developing countries, motorcycle ownership is predominantly among those in the middle and low income brackets \cite{Hagen2016}. This underscores the importance of accounting for the heterogeneity of transport users in our simulation model, reflecting common characteristics within groups and disparities between them, such as socioeconomic status. 
\\

After reviewing factors influencing mode choice previously identified in literature and those obtained with the multinomial models, we determined the variables to be included in the simulation model. Each agent is characterized by attributes such as sex, age, socioeconomic status, and a specific commuting distance, reflecting the population distribution in the city of Cali. During each decision period, agents calculate their trip satisfaction based on the selected mode. This calculation depends on their satisfaction with each mode attribute (acquisition cost, operating cost, comfort, road safety, personal security, travel time, and pollution) and the corresponding weights assigned to these attributes. The weights signify the importance agents place on each attribute. As indicated by the results of the multinomial models, these importance levels vary according to individual characteristics. Consequently, this variability underscores the need to determine the specific level of importance for each attribute for every agent in the simulation, a topic that will be further analyzed in the following subsection.

\subsection{Transport attributes perceptions and differences by users' groups}
\label{sec:3.2}
\begin{figure}[htb]
{
\centering
\includegraphics[width=0.35\textwidth]{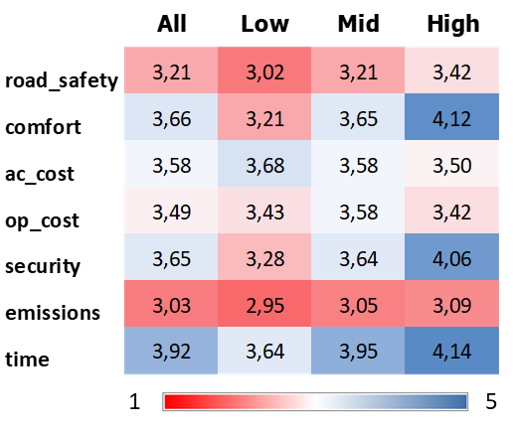}
\caption{Average importance level for transport attributes by socioeconomic group. Calculations based on the survey (ac.cost = acquisition cost, op.cost= operating cost).\label{fig:matrix}}
}
\end{figure}

Respondents rated the importance of each attribute using a Likert scale, where 5 represents the highest importance. Figure \ref{fig:matrix} presents the mean of attributes by socioeconomic groups. The Kruskal-Wallis test was employed to initially identify potential differences among the three socioeconomic groups: low, middle, and high. Based on the results displayed in Table \ref{tab:comparisons}, we can conclude that significant differences exist in the perceived relevance of road safety, personal security, operating costs, emissions, and travel time.
\\

Comparisons using the Chi-square ($\chi^2$) test reveal that there are no significant differences between the low- and middle-income groups regarding costs, road safety, and emissions. While the scores for these attributes are very similar, it should be noted that the importance ranking for these attributes differs between the groups. For low-income individuals, the three most important aspects of transport mode are acquisition cost, travel time, and operating cost, with personal security rated the lowest. In contrast, for the middle-income bracket, time, comfort, and personal security are deemed the most relevant. Pollution is considered the least important attribute by all groups, consistently receiving the lowest scores.
\\

\begin{table}[htb]
\centering
\caption{Chi-square ($\chi^2$) test results comparing the relevance of transport attributes by pairs of socioeconomic groups (***p-value \textless 0,01, **p-value \textless 0,05, *p-value \textless 0,1).\label{tab:comparisons}}
\begin{tabular}{|r|l|l|l|l|}
\hline
Attribute & Kruskal-Wallis & Low/Mid & Low/High & Mid/High \\ \hline
acquisition cost  & 0.76 & 0.17 & 0**   & 0.22 \\ \hline
operating cost    & 0*** & 0.41 & 0.24  & 0** \\ \hline
road safety       & 0*** & 0.11 & 0***  & 0**\\ \hline
personal security & 0*** & 0*** & 0***  & 0*** \\ \hline
comfort           & 0.35 & 0*** & 0***  & 0*** \\ \hline
time              & 0*** & 0**  & 0***  & 0** \\ \hline
emissions         & 0**  & 0.55 & 0.34  & 0.60 \\
\hline
\end{tabular}
\end{table}

Between the middle- and high-income groups, statistical differences are evident in the scores for all attributes except acquisition cost and emissions. However, these groups share the same order of priorities when evaluating transport modes, even though individuals at higher socioeconomic levels generally assign higher scores to all attributes.  The low and high levels have the greatest differences; the comfort offered by the transport mode is one of the most decisive things for wealthy users that can pay for it, whilst it is not relevant for the other group of users; also, crime and road accidents are not a concern for them. It can explain the high concentration of motorcycles--a low--cost and high--accident risk alternative \cite{CAF2015}---in low income classes that seek to maximize their budget at the expense of their security. 
\\

After identifying significant differences among the socioeconomic groups, the values were converted to scores between zero and one to be used as weights in the calculation of the satisfaction function. These scores are normally distributed among agents depending on their socioeconomic status.  

\subsection{Policy analysis}
\begin{figure}[htb]
{
\centering
\includegraphics[width=\textwidth]{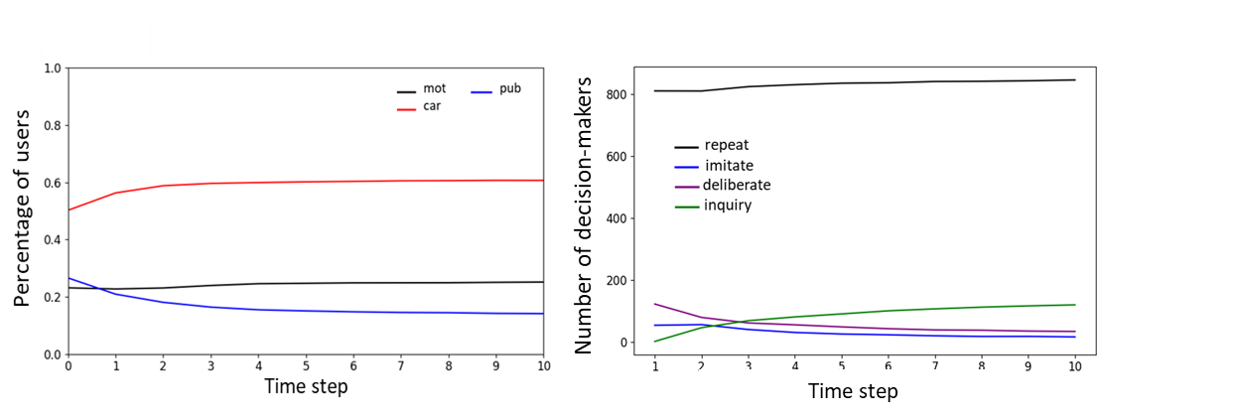}
\caption{(A) Simulation results for the base case scenario (mot = motorcycle, car= private car, pub = public transit). (B) Groups of decision-makers categorized by strategy employed.\label{fig:base-dm}}

}
\end{figure}

Using 2022 data from Cali to parameterize the model, we initialized it to simulate the distribution of transport users over the next decade (see Figure \ref{fig:base-dm}(A)). The projected decline in public transit usage aligns with current trends \cite{cali}. Without intervention, the model predicts that by 2032, only 14\% (C.I. 10\%-18\%) of commuters will use public transit--12\%less--, with significant increases in private transport usage, 22\% more car users and a 9\% increase in motorcycle usage, with participation of 61\% (C.I. 58\%-64\%) and  25\% (C.I. 21\%-29\%) respectively. These shifts are likely to worsen the city’s transportation issues, including an accident rate exceeding 11 per 100,000 people, increased CO2 emissions, and more unproductive hours spent in traffic jams. 
\\

Based on the transport attributes that agents evaluate after each commuting decision, we assess three public policy interventions aimed at preventing the migration from public to private modes of transport. Given that cost is a critical concern for low-income users, who comprise 35\% of the city’s population, we analyze the impact of implementing a free-fare policy for public transit—a strategy that has shown promising results in other contexts. The second strategy focuses on reducing travel time, the primary concern for middle- and high-income users, and the second concern for those in the lower income bracket. This is achieved by increasing the frequency of public transit services and reducing waiting times, thus improving overall travel times. The third policy targets personal security by taking measures to create a safer environment for transit users, which is expected to enhance their perceived safety towards this mode of transportation. 
\\

\begin{figure}[htb]
{
\centering
\includegraphics[width=1\textwidth]{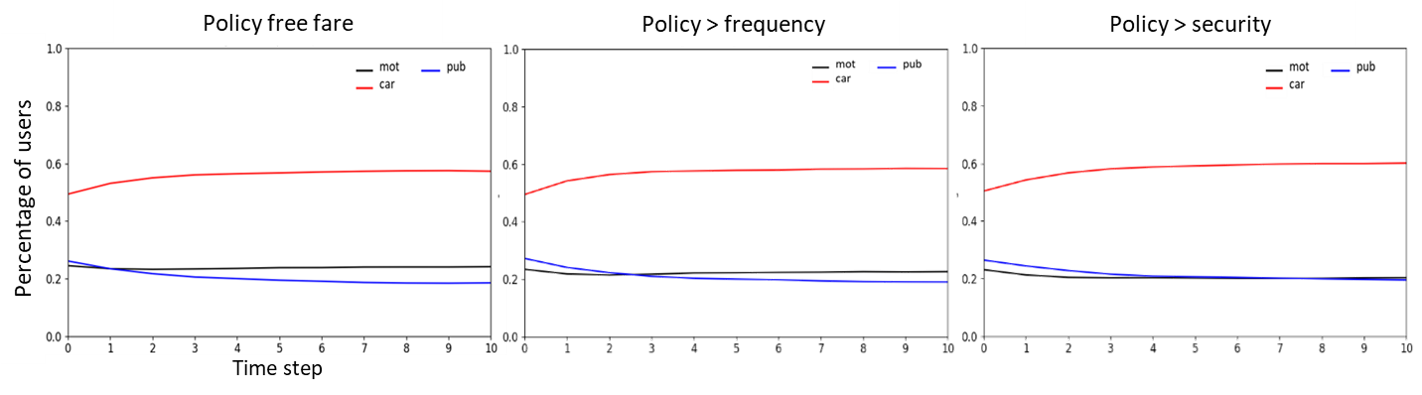}
\caption{Simulation results with the distribution of transport users after single policy intervention (mot = motorcycle, car= private car, pub = public transit).\label{fig:policies}}
}
\end{figure}

Figure \ref{fig:policies} presents the mean proportion of users per mode after the separate application of each policy. It is evident that all three policies positively impact the system by retaining more users in the public service than the base-case scenario. However, the net increase in users with any of the three policies implemented is only 5\% at end of the simulation period; this can be attributed to the observed inertia in the system where the majority of users experience low uncertainty and high satisfaction relative to their thresholds. As a result, they repeatedly choose the same transport modes. In contrast, a smaller proportion of users adopt behaviors guided by alternative strategies when faced with dissatisfaction and uncertainty (see Figure \ref{fig:base-dm}(B)).

\begin{figure}[htb]
{
\centering
\includegraphics[width=1\textwidth]{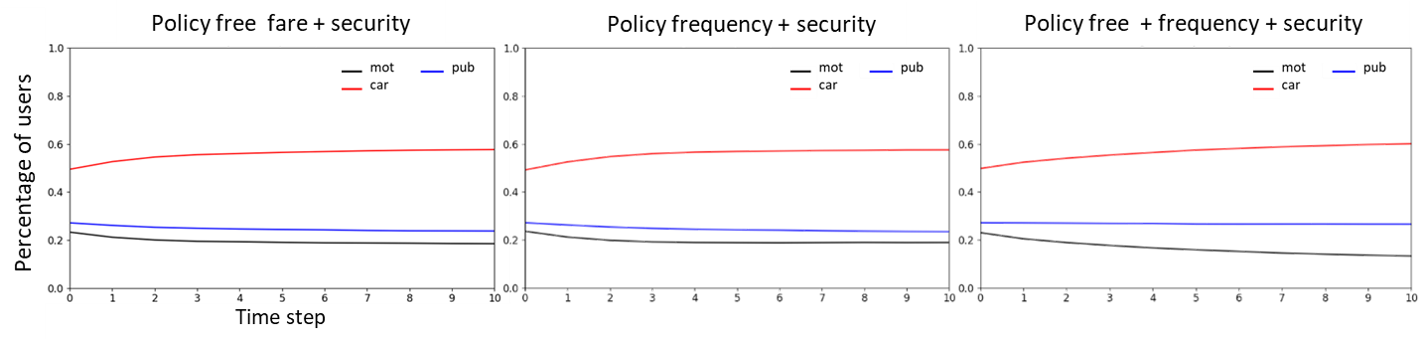}
\caption{Simulation results with the distribution of transport users after combined policy intervention (mot = motorcycle, car= private car, pub = public transit).\label{fig:combined}}
}
\end{figure}

In the second experiment, we analyze combinations of the policies, affecting two or three attributes simultaneously. The first combination—altering cost and personal security—presents a positive scenario, as shown in Figure \ref{fig:combined}. The participation of car users remains very similar to the base-case scenario (58\% vs. 61\%), but the proportion of motorcyclists decreases by 7\%, and public transit ridership increases by 10\%. When mixing frequency and security, the results are nearly identical. In contrast, the three-factor policy shows a more significant impact; the number of public transit users oscillates between 26.5\% and 27.2\% in the steps, while the proportion of motorcyclists decreases from 23\% to 13\%, and car users fluctuate between 50\% and 60\%. 
\\

In the model, decision-makers evaluate their satisfaction levels to decide whether to continue using the same transportation mode or to seek alternatives. Despite interventions on factors within the satisfaction function—a weighted sum of satisfaction for each transport attribute—the overall satisfaction values show minimal variation from the base case scenario. This can be explained by the similar ranges in attribute weights (all of them were ranked within the top half of the scale). Although certain attributes are more significant for specific user groups, the low dispersion of scores within these groups indicates that all attributes are important. Consequently, interventions on one single factor within the system have minimal impact on the satisfaction and subsequently on behavior. The values used for the weights represent socicultural aspects of the transport users and guide their behavior. In societies with characteristics and attribute relevance distribution similar to those studied in Cali, more effective results can be achieved by implementing policies that target multiple factors simultaneously; otherwise, efforts and resources directed at only one factor will be less effective.

\section{CONCLUSIONS}
This paper presents the results of a simulation model designed to represent mode choice decisions among urban commuters. The model serves as a testbed for evaluating the impacts of transport policies, aiming to contribute to the promotion of sustainable mobility. This endeavor represents a pioneering effort in developing an agent-based model tailored specifically for developing countries, where utilitarian travel heavily relies on motorcycles and exhibits distinct dynamics compared to those in the global North. By focusing on a Colombian city as a representative case, we introduced a methodology for identifying and incorporating sociocultural aspects that influence transport user behavior into simulation models. Utilizing multinomial models, we analyzed survey data collected from 970 respondents and conducted statistical tests to examine perceptions among transport users from different socioeconomic groups. This information was instrumental in customizing the model for policy analysis in developing countries.              
\\

Our analysis, conducted with the MNL and MNP models, underscores that urban mode choice is significantly influenced by sociodemographic factors such as age, gender, and socioeconomic status. These discrete choice models also highlight the importance of travel time and distance in mode selection. Furthermore  cost, road safety, personal security, time, comfort and travel time are variables significantly connected to travel decisions. Given the notable disparities between socioeconomic groups in developing countries, it was relevant to analyze the significance attributed to each of these influential attributes in decision-making; our findings reveal that cost and time are the most valued aspects for low-income users, while comfort and personal security hold greater importance for medium- and high-income individuals, surpassing cost considerations. Interestingly, pollution emerges as the least valued aspect across all groups, followed by road safety, despite the high accident rates in the region. These insights were integral to the development of our simulation model, wherein the importance levels assigned to these attributes were translated into weights used to calculate the overall satisfaction of commuters based on their socioeconomic status.           
\\

After identifying drivers of mode shift among travelers, representing them in the simulation model and validating results of a base case scenario, we tested three public policies. These interventions—free public transportation, doubled frequency in bus services, and enhanced security for system users—were positively received by current users, resulting in a reduced inclination towards private transportation alternatives. Compared to the base case scenario (without policies), the implemented policies led to a decrease in the number of motorcycles in circulation. However, each policy required a substantial investment in resources, resulting in only a 5\% increase in public transit users. Our findings suggest that policies that address multiple factors simultaneously are more effective than isolated actions, given that users consider all variables similarly relevant within this sociocultural context, which poses challenges for policymakers in devising cost-effective strategies.  
\\

Future work will involve comparing policy impacts using different settings of sociocultural variables. Additionally, further analysis is necessary to discern changes in travel behavior within specific socioeconomic groups, as the results presented in this paper focus on detailed user groups by mode. Moreover, a detailed discussion on the impacts of policies concerning CO2 emissions, travel time, and road accidents is needed.

\section*{ACKNOWLEDGMENTS}
The authors would like to express their gratitude to the Colombian Ministry of Science, Technology and Innovation (MinCiencias) for their financial support through the Technology and Innovation Program (CTeI). This funding made it possible to conduct the fieldwork necessary for applying the survey. GRANT: BPIN2022000100068.

\appendix

\section{SUPPLEMENTARY MATERIALS} \label{app:MN models}
A GitHub folder contains additional information including the Stata outputs for the multinomial models and the survey questionnaire. 
\url{https://github.com/Kathleenss/supplementary-WSC24}

\footnotesize

\bibliographystyle{wsc}

\bibliography{demobib}

\section*{AUTHOR BIOGRAPHIES}

\noindent {\bf \MakeUppercase{Kathleen Salazar-Serna}} is a PhD candidate in the Department of Computer and Decision Sciences at Universidad Nacional de Colombia - Medellín and an assistant professor at the School of Engineering and Sciences at Pontificia Universidad Javeriana in Cali. Her current research interests focus on sustainability issues and transport policy analysis. She uses agent-based modeling and network analysis to study transport dynamics. Her email address is   \email{kgsalaza@unal.edu.co} and her website is \url{https://orcid.org/0000-0003-3824-7044}.\\

\noindent {\bf \MakeUppercase{Lorena Cadavid}} is a professor in the Department of Computer and Decision Sciences at Universidad Nacional de Colombia Sede Medellín. In addition to her academic role, she is an enterprise consultant who applies her expertise to guide organizations towards data-driven decision making. Her research interest lies in policy design through modeling and simulation of social phenomena, and she leverages data analysis to support entrepreneurial decision making. Her email is \email{dlcadavi@unal.edu.co} and his website is \url{https://orcid.org/0000-0002-6025-5940}.\\

\noindent {\bf \MakeUppercase{Carlos J. Franco}} works as a full professor in the Department of Computer and Decision Sciences at Universidad Nacional de Colombia Sede Medellín. His research areas include complex systems, energy market modeling and simulation, and policy evaluation and strategy formulation. His email is  \email{cjfranco@unal.edu.co}.

\newpage

\end{document}